\documentclass[11pt,twoside]{article} 
\usepackage{newpasp,epsf}
\def\edcomment#1{\iffalse\marginpar{\raggedright\sl#1\/}\else\relax\fi} 
\reversemarginpar 

\begin{document} 
\title{Velocity Dispersion in the Halo of $\omega$ Centauri:
\hfil\break
No need for MOND given the present data}

\author{Dean E.~McLaughlin and Georges Meylan} 
\affil{STScI/ESA, 3700 San Martin Drive, Baltimore, MD 21218 USA}

\begin{abstract} 
We combine published data  to construct the radial-velocity dispersion
profile in $\omega$ Centauri over  projected radii $R\sim1.5$ to 60 pc
($\sim$0.4--15 core  radii). All sources  of data are  consistent with
each other and are well described at large radii by standard dynamical
models; they do not require modified Newtonian dynamics (MOND) or dark
matter.
\end{abstract}

Scarpa, Marconi,  \& Gilmozzi (2003) have used  their new measurements
of  the  line-of-sight  velocity  dispersion at  $R\sim30$--45  pc  in
$\omega$ Centauri,  to argue  for modifications to  Newtonian dynamics
(MOND: Milgrom 1983) at low accelerations in the outer reaches of this
cluster.   Alternatively,  any such  evidence  could  be  taken as  an
indication of  the presence of  an extended dark-matter halo.  Here we
show  that the  data  of Scarpa  et  al.~are in  fact consistent  with
earlier observations  by Seitzer (1983)  in the outskirts  of $\omega$
Cen,  which are  in  turn well  described  by simple,  self-consistent
dynamical models.

The  top  panels of  Figure  1  show  the $V$-band  surface-brightness
profile of  $\omega$ Cen, taken  from Meylan (1987) and  corrected for
reddening.  A  heliocentric distance of  5.1 kpc is assumed.  The {\it
dashed curves}  trace the best  fit of an isotropic,  single-mass King
(1966) model,   in  which   the  stellar   distribution   function  is
$f(E)\propto [\exp(-E/\sigma_0^2)-1]$ for $E<0$.  The poor fit of this
model  around  the tidal  radius  has been  remedied  in  the past  by
allowing for velocity anisotropy  and for energy equipartition between
multiple  mass classes  (Meylan 1987;  Meylan et  al.~1995).  However,
observations now show  that the velocity ellipsoid in  this cluster is
everywhere isotropic  (Merritt et  al.~1997; van Leeuwen  et al.~2000;
King \& Anderson 2002), and that mass segregation is not as pronounced
as expected for equipartition (Anderson  1997). Thus, we revert to the
assumptions  of  velocity isotropy  and  a  single  stellar mass,  and
instead obtain an improved fit to the density profile with a different
model     for      the     distribution     function:     $f(E)\propto
[\exp(-E/\sigma_0^2)+E/\sigma_0^2-1]$  for  $E<0$  (Wilson 1975;  {\it
solid curves} in Fig.~1).  \looseness=-2

The bottom  panels of Figure  1 show the  observed velocity-dispersion
profile in $\omega$ Cen over projected radii $R\simeq 1.5$--60 pc. The
hatched region, defined by $30\le  R\le 45$ pc and $6.5\le \sigma_p\le
9.5$  km s$^{-1}$,  represents  the range  of  the data  of Scarpa  et
al.~(2003). They  are consistent with the other  measurements at these
radii. Curves  show the projected $\sigma_p(R)$  profiles predicted by
the single-mass, isotropic Wilson (1975) and King (1966) models fit to
the stellar density distribution in  the upper panels, normalized by a
reasonable,    spatially    constant    mass-to-light    ratio.    All
velocity-dispersion data  in the halo are fully  consistent with these
models.   The   brief   ``flattening''   of   $\sigma_p(R)$   at   the
clustercentric distances  observed by Scarpa et  al.~(2003) would have
to extend much farther to be taken as positive support either for MOND
or for the presence of dark matter.

\begin{figure}[t]
\plotfiddle{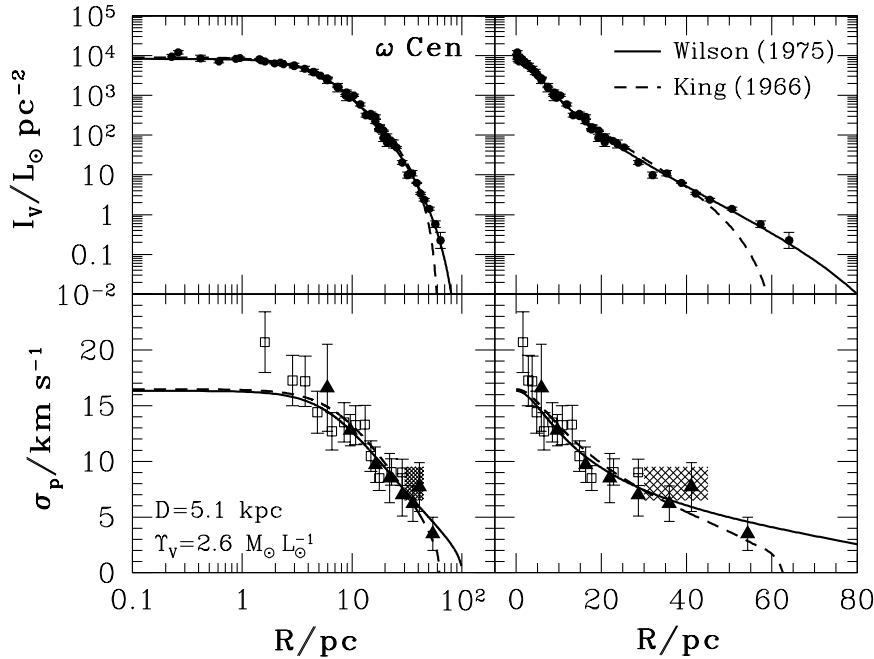}{3.2truein}{0}{60}{60}{-180}{-130}
\caption{Comparison  of the  radial-velocity  dispersions measured  by
Scarpa  et  al.  (2003;  hatched  regions) with  those  of  Meylan  et
al.~(1995;  open squares)  and Seitzer  (1983; filled  triangles). The
best-fit self-consistent  models of Wilson (1975) and  King (1966) are
shown; see text.  Neither MOND  nor dark matter is required by current
data in the halo of $\omega$ Centauri.}
\end{figure}

\end{document}